\documentclass[journal,]{IEEEtran}

\usepackage{enumerate}
\usepackage{cite}
\usepackage{tikz}
\usepackage{xcolor}
\graphicspath{{figures/}}
\usepackage{algorithm }
\usepackage{algorithmic}

\usepackage{makecell}
\usepackage{bm}
\usepackage{amsmath,amsfonts,amssymb,amsthm}
\usepackage{subfigure}
\usepackage{threeparttable}
\usepackage{booktabs}
\usepackage{bm}
\usepackage{mathrsfs}

\usepackage{url}

\makeatletter
\def\UrlAlphabet{%
      \do\a\do\b\do\c\do\d\do\e\do\f\do\g\do\h\do\i\do\j%
      \do\k\do\l\do\m\do\n\do\o\do\p\do\q\do\r\do\s\do\t%
      \do\u\do\v\do\w\do\x\do\y\do\z\do\A\do\B\do\C\do\D%
      \do\E\do\F\do\G\do\H\do\I\do\J\do\K\do\L\do\M\do\N%
      \do\O\do\P\do\Q\do\R\do\S\do\T\do\U\do\V\do\W\do\X%
      \do\Y\do\Z}
\def\UrlDigits{\do\1\do\2\do\3\do\4\do\5\do\6\do\7\do\8\do\9\do\0}
\g@addto@macro{\UrlBreaks}{\UrlOrds}
\g@addto@macro{\UrlBreaks}{\UrlAlphabet}
\g@addto@macro{\UrlBreaks}{\UrlDigits}
\makeatother

\usepackage{footnote}
\makesavenoteenv{tabular}

\begin{document}

\title{Inertial Sensing Meets Artificial Intelligence: Opportunity or Challenge? }
\author{You Li, 
  Ruizhi Chen, 
  Xiaoji Niu,  
  Yuan Zhuang,
  Zhouzheng Gao, 
  Xin Hu,
  and Naser El-Sheimy
   \thanks{Y. Li and N. El-Sheimy are with the University of Calgary (liyou331@gmail.com; elsheimy@ucalgary.ca). R. Chen, Y. Zhuang, and X. Niu are Wuhan University (ruizhi.chen@whu.edu.cn; zhy.0908@gmail.com; xjniu@whu.edu.cn). Z. Gao is with China University of Geosciences (Beijing) (zhouzhenggao@126.com). X. Hu is with the School of Electronic Engineering, Beijing University of Posts and Telecommunications (huxin2016@bupt.edu.cn).  
   }
}

\markboth{In preparation}%
         {Shell \MakeLowercase{\textit{et al.}}: Bare Demo of IEEEtran.cls for Journals}

         \maketitle

         \begin{abstract}
The inertial navigation system (INS) has been widely used to provide self-contained and continuous motion estimation in intelligent transportation systems. Recently, the emergence of chip-level inertial sensors has expanded the relevant applications from positioning, navigation, and mobile mapping to location-based services, unmanned systems, and transportation big data. Meanwhile, benefit from the emergence of big data and the improvement of algorithms and computing power, artificial intelligence (AI) has become a consensus tool that has been successfully applied in various fields. This article reviews the research on using AI technology to enhance inertial sensing from various aspects, including sensor design and selection, calibration and error modeling, navigation and motion-sensing algorithms, multi-sensor information fusion, system evaluation, and practical application. Based on the over 30 representative articles selected from the nearly 300 related publications, this article summarizes the state of the art, advantages, and challenges on each aspect. Finally, it summarizes nine advantages and nine challenges of AI-enhanced inertial sensing and then points out future research directions. 
\end{abstract}


         \IEEEpeerreviewmaketitle

\section{Introduction}  
\label{sec-intro}
         \IEEEPARstart{T}{he} inertial navigation system (INS) has been widely used in the field of intelligent transportation systems due to its capability to provide self-contained and continuous motion information \cite{Barbour-Schmidt-2001}. In recent years, with the development of micro-electro-mechanical (MEMS) technology, chip-level inertial sensors (i.e., gyros and accelerometers) with low cost and low power consumption have appeared. Such sensors have expanded related applications from inertial navigation to inertial sensing, that is, from traditional positioning, navigation, and mobile mapping to location-based services, smart wearables, autonomous systems, motion big data, inertial surveying, and many other emerging fields \cite{ElSheimy-Youssef-2020}.

Meanwhile, because of the emergence of big data and the improvement of algorithms and computing power, artificial intelligence (AI) has become a powerful tool that has been successfully applied in many industries \cite{LeCun-Bengio-2015}. In the field of positioning, navigation, and location-based services, AI technology has been widely used in image processing, point-cloud processing, and motion-data analysis. The introduction of AI has solved some industry challenges and even fundamentally changed the way of processing and using data \cite{Li-D-2018}. AI technology has also been used in positioning and navigation based on wireless signals (e.g., global navigation satellite systems (GNSS), WiFi, and 5G cellular) \cite{Lin-Li-2020} and environmental (e.g., magnetic, sound, and point cloud) features \cite{Hata-Ramos-2018}. In comparison, the research works on AI-based inertial data processing are less; however, the number of studies is increasing.

        This article reviews the basic ideas and main methods of using AI technology to enhance inertial sensing. Furthermore, it summarizes and analyzes the research status, advantages, challenges, and trends of related technologies. This article is organized as follows: Section \ref{sec-ai-enhanced-inertial-sensing} reviews the research on AI-enhanced inertial sensing. Section \ref{sec-AI-algorithms} summarizes the AI algorithms used. Section \ref{sec-pros-and-cons} analyzes the advantages and challenges of using AI for inertial sensing. Finally, Section \ref{sec-conclusion} is a summary and outlook.
     
     \section{Artificial-Intelligence-Enhanced Inertial Sensing}
\label{sec-ai-enhanced-inertial-sensing}
   
\begin{figure}
           \centering
           \includegraphics[width=0.38 \textwidth]{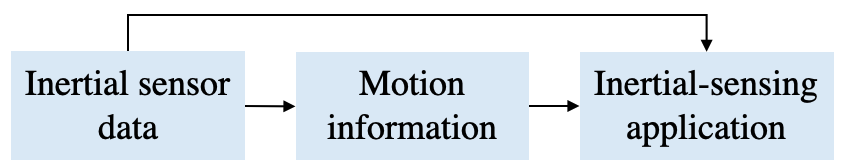}
           \caption{Relationship between inertial-sensing application, inertial sensor data, and motion information}
           \label{fig:relation}
         \end{figure}

\begin{figure}
           \centering
           \includegraphics[width=0.45 \textwidth]{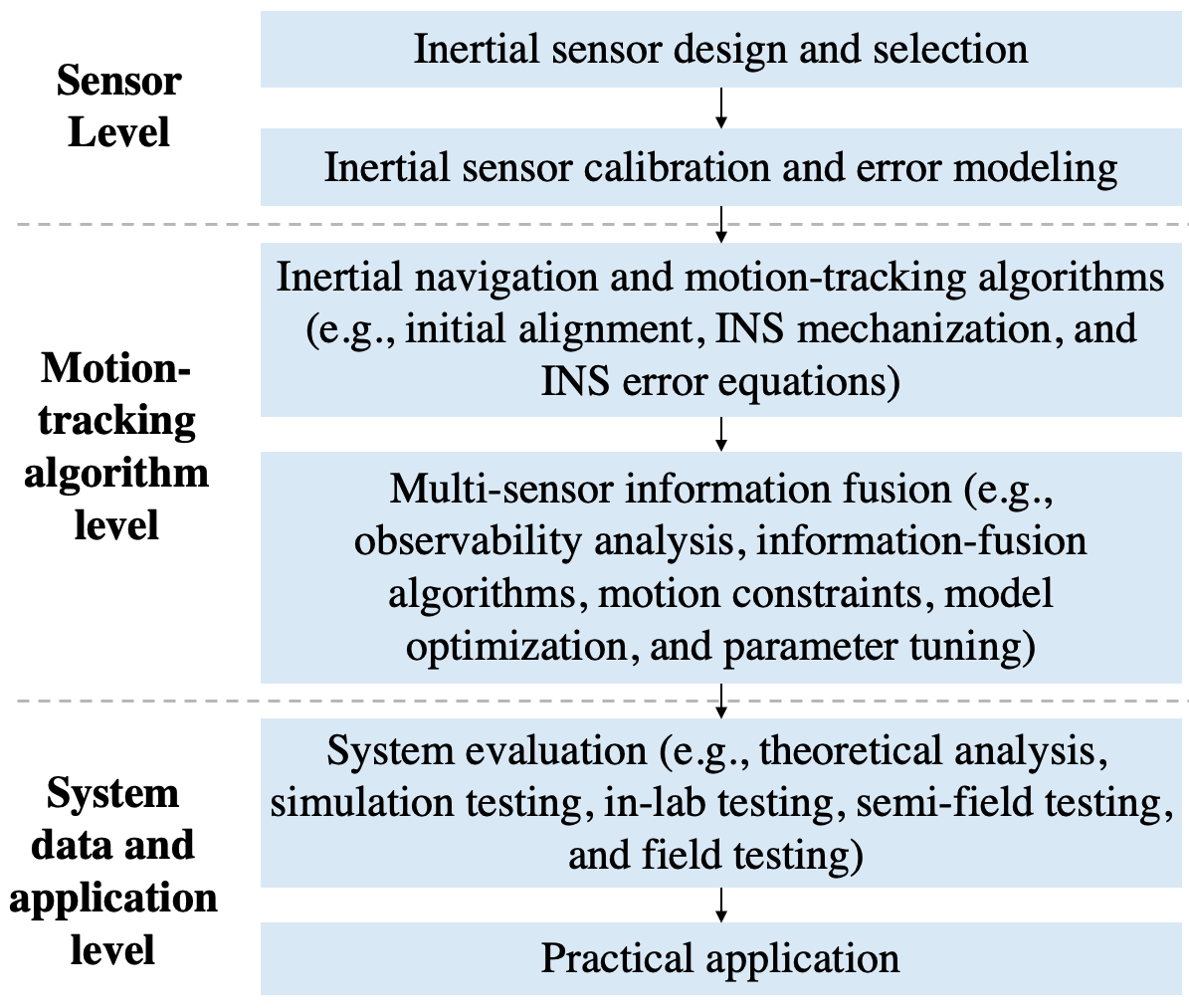}
           \caption{Main steps in inertial-sensing applications}
           \label{fig:steps}
         \end{figure}

        Compared to traditional inertial navigation, which determines motion information based on the inertial sensor data, inertial sensing has a wider range. As shown in Figure \ref{fig:relation}, it includes not only inertial navigation but also the sensing applications based on either motion information or sensor data. The main steps of inertial-sensing activities can be summarized as sensor design and selection, sensor calibration and error modeling, navigation and motion-tracking algorithms, multi-source information fusion, testing and evaluation, and practical applications, as shown in Figure \ref{fig:steps}.   

These steps affect inertial-sensing performance from different levels. To achieve specific sensing accuracy and cost requirements, proper sensor design and selection are essential. Calibration and error modeling improve measurement accuracy from the sensor level. Meanwhile, better algorithms of inertial navigation, motion tracking, and multi-source information fusion can enhance the sensing performance from the algorithm level. In contrast, testing, evaluation, and practical applications are carried out from the system data and application level. To the best of the authors’ knowledge, this article is the first systematic review of AI-enhanced inertial sensing, involving all steps in Figure \ref{fig:steps}.

According to the key factors such as the role in inertial sensing, AI model, and application platform, over 30 representative articles are selected from the nearly 300 related publications. The selected research works are listed in Table \ref{tab-literature}. Also, they are introduced in the following subsections.

\begin{table*}
           \centering
\begin{tabular}{c p{3.6cm} p{2.2cm} p{2.8cm} p{2.8cm} p{1.6cm} p{1.5cm}}
\hline
\textbf{Ref.} & \textbf{Aspect in inertial sensing}  &  \textbf{AI model type}  &  \textbf{AI input}  &  \textbf{AI output} &  \textbf{Platform} & \textbf{Sensor} \\ \hline
\cite{Hsu-C-2011} & Sensor hardware &  ANN (ANFIS)	  &  Control input	& Control output  &	Unspecified &	Inertial \\ 
\cite{Zha-Hu-2009} & Sensor hardware &  ANN (BP) & Start and stop time & Engine state &	Unspecified	& Inertial \\ 
\cite{Chen-X-2003} & Calibration and error modeling	& ANN (BP)	& Sensor data	& Estimated bias 	& Unspecified	& Inertial \\ 
\cite{Chong-S-2016} & Calibration and error modeling	& Genetic algorithm	& Sensor data, thermal data	& Estimated thermal drift 	& Unspecified	& Inertial \\ 
\cite{Nobre-Heckman-2019}& Calibration and error modeling	& DRL	& Sensor data	& Motion type	& Robot	& Inertial, vision \\ 
\cite{Kim-Agrawal-2016}&  Calibration and error modeling	& Empirical mode decomposition 	& Sensor data	& Reconstructed data	 & Wearable	& Inertial \\ 
\cite{Chiang-Huang-2010} & Inertial navigation and motion-tracking algorithm	& ANN (ANFIS)& 	Attitude	& Attitude error& 	Car	& Inertial \\ 
\cite{Jaradat-Abdel-2014}& Inertial navigation and motion-tracking algorithm	& ANN (ANFIS)	& Motion state	& Motion state error& 	Car	& Inertial \\ 
\cite{Indelman-Williams-2013} &Inertial navigation and motion-tracking algorithm	& Factor graph	& Motion state, sensor data	& Motion state	& Car	& Inertial \\ 
\cite{Chen-Zhao-2020} &Inertial navigation and motion-tracking algorithm & 	ANN (LSTM)& 	Sensor data	& Motion state& 	Wearable	& Inertial \\ 
\cite{Wagstaff-Kelly-2018}& Multi-sensor fusion algorithm& 	ANN (LSTM)& 	Sensor data& 	Zero-velocity detection& 	Shoe	& Inertial \\ 
\cite{Gonzalez-Fiacchini-2018} & Multi-sensor fusion algorithm& 	GP	& Sensor data	& Sideslip detection	& Robot	& Inertial \\ 
\cite{Tong-X-2018}& Multi-sensor fusion algorithm	& HMM	& Sensor data	& Magnetic interference detection& 	Wearable & 	Inertial, magnetic \\ 
\cite{Brossard-Barrau-2020}& Multi-sensor fusion algorithm	& ANN (CNN)	& Sensor data& 	Motion uncertainty prediction& 	Car& 	Inertial \\ 
\cite{Wu-Yang-2010}& Multi-sensor fusion algorithm	& Adaptive Kalman filter	& Motion state, sensor data	& Weight of information	& Car& 	Inertial, GNSS \\ 
\cite{Milford-M-2013}& Multi-sensor fusion algorithm& 	Brain-like algorithm& 	Sensor data	& Navigation correction& 	Drone	& Inertial, vision \\ 
\cite{Li-Wang-2014}& Multi-sensor fusion algorithm	& ANN (ANFIS)	& Motion state	& Navigation correction	& Car& 	Inertial, odometer, GNSS \\ 
\cite{Rambach-Tewari-2016}& Multi-sensor fusion algorithm& 	ANN (LSTM)	& Motion state, sensor data	& Navigation correction& 	Augmented reality& 	Inertial, vision \\ 
\cite{Buskey-Wyeth-2001}& System data and application& 	ANN (BP)	& Motion state, sensor data	& Motion control	& Helicopter& 	Inertial \\ 
\cite{Ravi-Wong-2016}& System data and application& 	ANN (CNN)	& Sensor data	& Motion mode& 	Wearable& 	Inertial \\ 
\cite{Moschetti-Fiorini-2016}& System data and application& 	SVM, random forests	& Sensor data	& Activity mode	& Ring& 	Inertial \\ 
\cite{Eskofier-B-2016}& System data and application	& ANN (CNN)	& Sensor data	& Disease mode	& Wearable	& Inertial \\ 
\cite{Khabir-K-2019}& System data and application& 	SVM, decision trees	& Step data	& Age and gender	& Wrist	& Inertial \\ 
\cite{Kim-Cho-2020}& System data and application& 	ANN (LSTM)& 	Sensor data& 	Motion mode	& 17 wearables& 	Inertial \\ 
\cite{Windau-Itti-2019}& System data and application	& ANN (RNN)	& Sensor data& 	Motion feedback	& 13 wearables& 	Inertial \\ 
\cite{Yang-Ahn-2016}& System data and application	& SVM	& Sensor data	& Safety mode	& Sacrum	& Inertial \\ 
\cite{Li-Xie-2019}& System data and application& 	ANN (CNN)& 	Sensor data& 	Driving habit mode& 	Wrist	& Inertial \\ 
\cite{Allouch-Kouba-2017}& System data and application	& Decision trees	& Sensor data	& Road condition	& Smartphone on car& 	Inertial \\ 
\cite{Gonzalez-Hidalgo-2008}& System data and application& Unspecified	& Motion data& 	Spatiotemporal analysis& 	Smartphone& 	Smartphone sensors \\ 
\cite{Han-Liang-2015}& System data and application	& SVM	& Motion data	& Trend prediction& 	Smartphone	& Smartphone sensors \\ 
\cite{Li-Cai-2016}& System data and application	& ANN (wavelet)	& Motion state	& Data weight	& Pipeline	& Inertial \\ 
\cite{Lertxundi-Diez-2008}& System data and application& 	ANN (general)	& Sensor data	& Railway mode	& Railway	& Inertial \\ 
\cite{Li-X-2009}& System data and application	& ANN (general)	& Motion state	& Gravity disturbance	& Car& 	Inertial, GNSS \\ 
 \hline
           \end{tabular}
           \begin{tablenotes}
        		\item[*] ~~~* ANFIS - adaptive neuro-fuzzy inference system; ANN - artificial neural network; DRL - deep reinforcement learning; LSTM - long short-term memory; CNN - convolutional neural network; RNN - recurrent neural network; SVM - support vector machine; GP - Gaussian process
     		\end{tablenotes}
           \caption{ Selected research on AI-enhanced inertial sensing   }
           \label{tab-literature}
         \end{table*}
         
\subsection{Artificial-Intelligence-Enhanced Inertial-Sensor Hardware}

The inertial system calculates the changes of motion states (e.g., position, velocity, and attitude) by integrating the gyro-derived angular velocities and accelerometer-derived specific forces. The computed motion changes at each time epoch are added into the previous motion state to obtain real-time dead-reckoning solutions. Therefore, the inertial sensor hardware is the basis of inertial navigation and inertial sensing \cite{Titterton-Weston-2004}. 

In terms of inertial sensor hardware, the existing literature uses AI technology for gyro control and gyro-life prediction. The principle of AI-enhanced gyro control \cite{Hsu-C-2011} is to use the back-propagation (BP) characteristics of artificial neural network (ANN, including deep neural networks (DNN)) for online estimation of the system dynamic function. In contrast, the idea of using AI for gyro-life prediction \cite{Zha-Hu-2009} is to use AI technology to establish the relationship between the state of the gyro motor and the gyro’s start and stop times. Specifically, the research work \cite{Zha-Hu-2009} proposes a hybrid model that applies a BP-ANN to predict the correction of a gray model, so as to obtain time-series analysis results that are superior to both the gray model and BP-ANN. 

This research reflects the idea of using AI instead of complex models for time-series analysis, especially when the model is difficult to describe explicitly. However, it simply compares the prediction errors of three models but has not discussed the internal structure of ANN and its relationship with time-series analysis in depth. Therefore, it has not quantified the specific impact of ANN on the time-series analysis system. 

In general, there are limited research works on using AI technology to enhance inertial-sensor hardware. However, it is expected that with the development of AI-hardware technology \cite{LeCun-Y-2019}, AI-enhanced inertial sensors may become an important research direction.

\subsection{Artificial-Intelligence-Enhanced Inertial-Sensor Calibration and Error Modeling}

          \begin{figure}
           \centering
           \includegraphics[width=0.44 \textwidth]{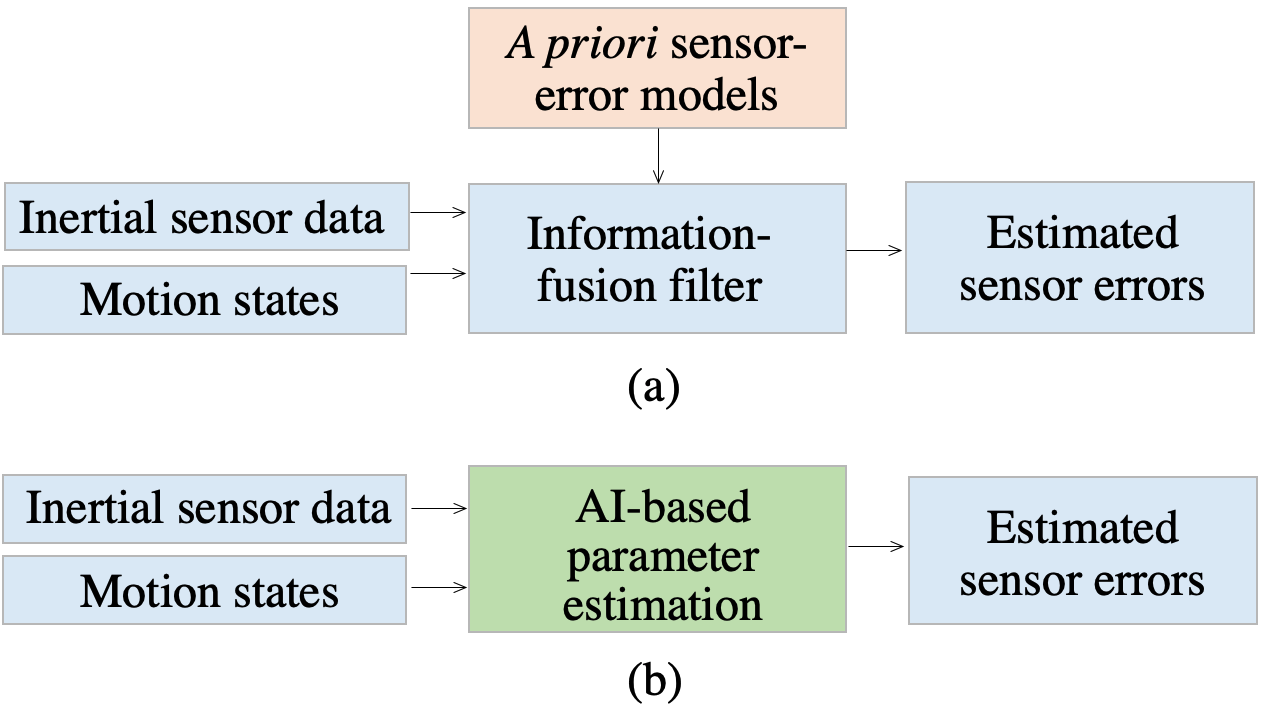}
           \caption{Inertial sensor error estimation using traditional (a) and AI (b) methods}
           \label{fig:module-err-est}
         \end{figure}   
         
Inertial sensor errors are an important error source in inertial sensing. Uncompensated sensor errors accumulate as motion state errors over time. In general, a gyro bias causes an attitude, velocity, and position error that increases with the first, second, and third power of time, respectively. Also, an accelerometer bias causes a velocity and position error that increases with the first and second power of time, respectively \cite{Li-Y-Thesis-2016}. Therefore, it is essential to control inertial sensor errors. 

Inertial sensor errors can be divided into deterministic errors (e.g., biases, scale-factor errors, and non-orthogonal errors) and stochastic errors (e.g., instabilities and noises). Deterministic sensor errors can be determined by calibration. However, laboratory calibration is costly and too ideal to reflect the actual application environment. Therefore, low-cost applications often require online sensor-error estimation \cite{Li-Y-Cali-2015}. 

To estimate inertial sensor errors, AI technology has been applied. Figure \ref{fig:module-err-est} illustrates the principle of AI-based inertial sensor-error estimation, which uses AI to replace the traditional approach based on \textit{a priori} sensor-error models and an information-fusion filter. The literature \cite{Chen-X-2003} finds that ANN can obtain a more accurate gyro bias estimation than the traditional method based on the combination of autoregressive moving average (ARMA) model and Kalman filter. Also, ANN does not require \textit{a priori} information on the sensor noise level. Meanwhile, the research work \cite{Chong-S-2016} shows that ANN can obtain more accurate gyro thermal-drift estimation than polynomial fitting. Using these two thermal-drift compensation models reduces the inertial navigation error by 81 \% and 49 \%, respectively. 

On the other hand, it has been pointed out that compared to traditional methods such as Fourier analysis, wavelet transform, and Kalman filtering, AI methods lack the basis of mathematical models and thus they rely more on training data. The research works \cite{Chen-X-2003} and \cite{Chong-S-2016} have not examined the performance of ANN when the training data is insufficient or the training and testing data are collected in various scenarios. Also, they have not explored the impact of AI structure and parameters, which directly affect the performance of parameter estimation.

\begin{figure}
           \centering
           \includegraphics[width=0.38 \textwidth]{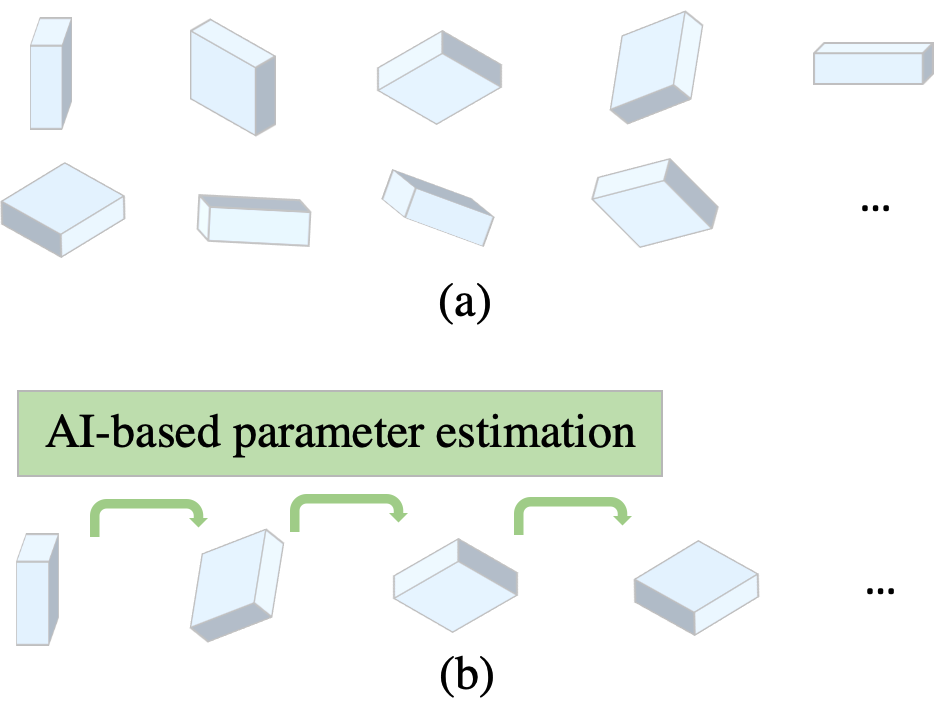}
           \caption{Inertial sensor calibration operation without (a) and with guide from AI (b)}
           \label{fig:module-cali}
         \end{figure}

In terms of sensor calibration, the research in \cite{Nobre-Heckman-2019} uses reinforcement learning to optimize calibration actions and improve calibration accuracy and efficiency. Figure \ref{fig:module-cali} shows the principle of AI-aided inertial sensor calibration, which applied AI to determine the sequence of calibration operation in real time. Compared with the geometric action-optimization method, the reinforcement-learning-based method obtains a calibration error with a similar mean and smaller standard deviation. More importantly, the reinforcement-learning-based method can be completed by people without professional visual-inertial calibration experience. The AI method used in this research may also enhance other decision-making-based inertial-sensing applications, such as mobile-mapping trajectory planning and adjustment. 

The use of AI can reduce manual intervention and help achieve autonomous systems. However, the test data in \cite{Nobre-Heckman-2019} is similar to the training scenario; thus, the reinforcement-learning results may be affected by overfitting. In addition, the reinforcement learning method is data-driven; thus, its inside algorithm is more difficult to physically and geometrically explain compared to traditional methods such as observability analysis. This factor brings a challenge for a deeper understanding of the problem.
         
In addition, because inertial systems continuously calculate motion states based on inertial sensor data, the continuity of the data is crucial. The research work \cite{Kim-Agrawal-2016} presents an AI-based data-reconstruction method that can recover inertial sensor data lost during wireless transmission. The results show that its new method, which combines the autoregressive (AR) model and empirical mode decomposition, results in inertial sensor data reconstruction results that are superior to both models. This phenomenon indicates that the combination of AI technique and traditional time-series analysis method may obtain better performance than the two. The method in this study is applicable to not only inertial data but also other time series, such as GNSS data, clock data, and navigation results.  

\subsection{Artificial-Intelligence-Enhanced Inertial-Navigation and Motion-Tracking Algorithms}

The inertial navigation algorithm inputs inertial sensor data and outputs motion states. The algorithm is divided into two stages: initial alignment and dead-reckoning. Initial alignment provides an initial position and attitude, which are essential for dead-reckoning \cite{Shin-E-Thesis-2005}. The research in \cite{Chiang-Huang-2010} applies an adaptive neuro-fuzzy inference system (ANFIS) to predict attitude errors, so as to alleviate the slow convergence of heading estimation in the traditional initial-alignment Kalman filter. The introduction of ANN respectively reduced the initial-heading error and alignment time by 66 \% and 50 \% when a high-end tactical IMU was used, and both over 85 \% when a navigation-level IMU was used. 

This study shows the potential of using ANNs to enhance initial alignment. It provides an idea to use AI to accelerate the convergence of Kalman filtering. This idea is applicable to not only initial alignment but also other Kalman-filter-based engineering problems, such as real-time navigation. On the other hand, this research cannot guarantee how much performance will be improved because the training and test data were collected in the same scenario. In addition, the study did not reveal the relationship between ANN parameters and initial alignment performance. 

\begin{figure}
           \centering
           \includegraphics[width=0.46 \textwidth]{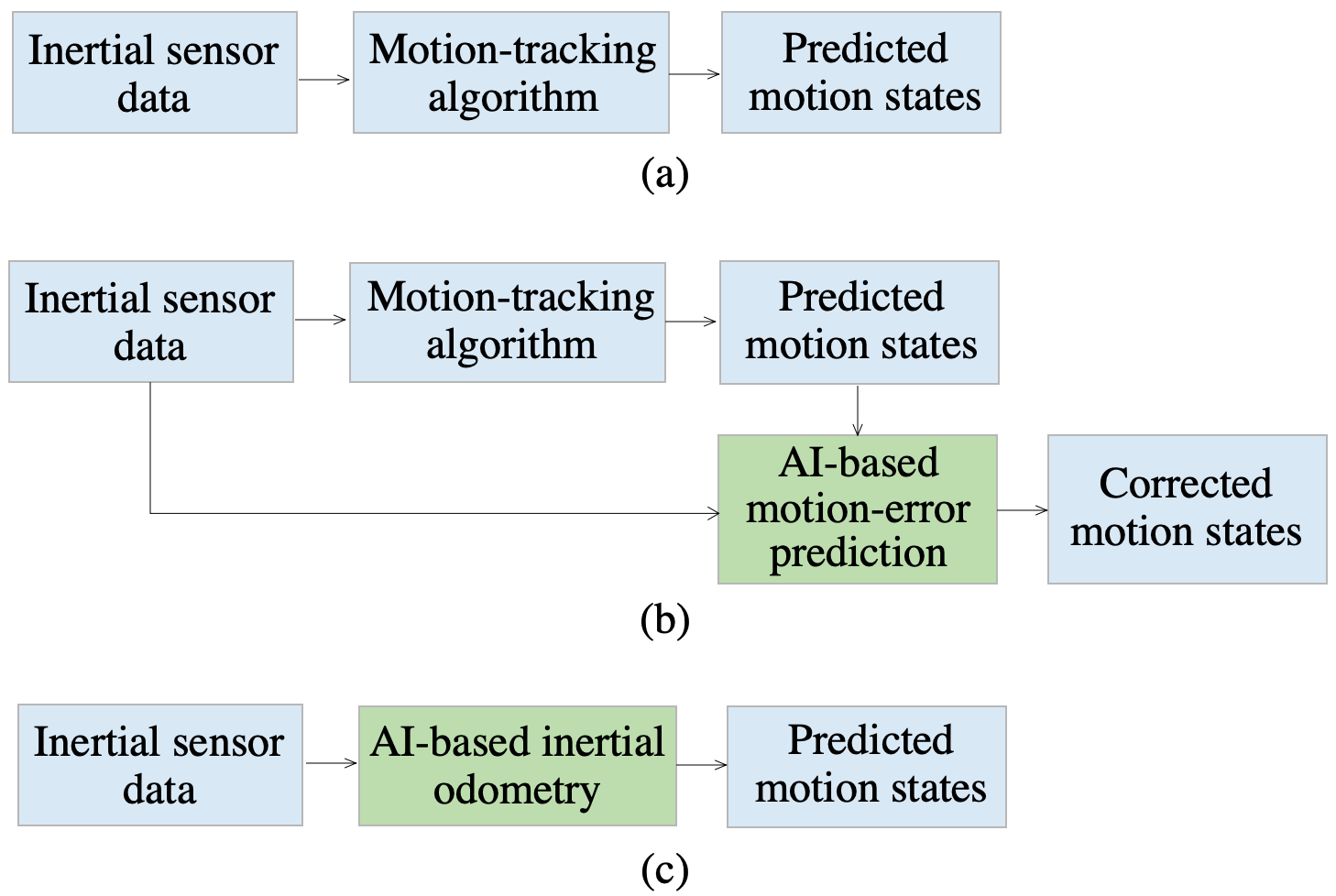}
           \caption{Inertial sensor based motion tracking with various strategies: (a) traditional dead-reckoning algorithm, (b) dead-reckoning algorithm with AI-based correction; (c) AI-based inertial odometry}
           \label{fig:module-motion-tracking}
         \end{figure}

Dead-reckoning is mainly realized by using geometric algorithms, such as three-dimensional (3D) INS mechanization and 2D pedestrian dead-reckoning (PDR). The research work \cite{Jaradat-Abdel-2014}, which adopts a similar idea as the paper \cite{Chiang-Huang-2010}, applies ANFIS to predict INS navigation error correction, so as to maintain its navigation accuracy. Figure \ref{fig:module-motion-tracking} (b) shows the principle. In contrast, the research in \cite{Indelman-Williams-2013} uses factor graphs to smooth navigation results and improve the sensing accuracy. To be specific, it attempts to apply incremental smoothing to achieve a performance that is close to batch processing. Different from traditional smoothing approaches, this method can achieve quasi-real-time processing; however, it is still difficult to apply in real time. 

These studies show that AI can be used to enhance not only real-time parameter estimation but also post-processing estimation. Furthermore, the introduction of AI can not only improve parameter-estimation but also reduce manual intervention for parameter tuning. A challenge for these research works is that they may need a complex AI model and a large amount of training data. Meanwhile, they have not revealed the relationship between the AI model and estimation performance. Furthermore, they have not compared AI-based parameter tuning to other approaches, such as manual tuning.

In most research works, the AI-enhanced inertial motion tracking can be regarded as a combination of traditional geometric algorithms and AI algorithms. Among them, the geometric algorithm builds the mathematical model from IMU data to motion states, while the AI algorithm is used to enhance or replace the complex modules in the geometric method. 

Different from these works, literature \cite{Chen-Zhao-2020} attempts to use AI to completely replace geometric algorithms. To be specific, it utilizes an ANN-based inertial-odometry algorithm, instead of traditional INS mechanization and PDR algorithms, to directly generate motion states from IMU data, as shown in Figure \ref{fig:module-motion-tracking} (c). The advantage is that it can mitigate the influence of some error factors that are difficult to model in geometric algorithms. Examples of these error sources are the initial-alignment error and the misalignment angle between the device and vehicle platform. From the results, The ANN algorithm significantly mitigated the position drifts on a large spatial scale. One possible reason for this improvement is that the ANN-based approach does not contain an integral calculation step, which exists in the INS and PDR algorithms. 

On the other hand, when using the ANN-based inertial-odometry algorithm to completely replace the geometric algorithm, the navigation system lost an important advantage of inertial navigation: the short-term high accuracy and reliability. From the results, the ANN-based navigation results fluctuate and deform on a small spatial scale. For example, when the actual navigation trajectory is square, the calculated trajectory is petal-shaped. 

Considering the above advantages and disadvantages of the ANN-based inertial-odometry algorithm, it is recommended to fuse it with the geometric algorithm properly. For a profound fusion, how to determine the weights of geometric and AI algorithms in different scenarios is a challenge. Also, the research in \cite{Chen-Zhao-2020} has neither evaluated the performance of the ANN-based approach in different scenarios nor given quantitative evaluation results. Meanwhile, it lacks a theoretical guide for the choice of ANN type, model layers, and the number of neurons. These factors also need to be considered for further research.

\subsection{Artificial-Intelligence-Enhanced Multi-Sensor Information Fusion}

Because inertial motion-tracking solutions drift over time, external constraints are needed to correct the results. The commonly-used constraints are from hardware sensors and algorithm constraints \cite{Skog-Handel-2009}. The traditional approaches to determine whether a constraint is needed include field testing and theoretical-analysis methods such as observability analysis. 

In recent years, AI technology has been used to detect and use motion constraints, such as zero-velocity updates \cite{Wagstaff-Kelly-2018} and non-holonomic constraints \cite{Gonzalez-Fiacchini-2018}, and environmental constraints, such as the interference of magnetic fields \cite{Tong-X-2018}. In the paper \cite{Wagstaff-Kelly-2018}, ANN is applied to replace the traditional zero-velocity detection method, which is based on analysis of data features (e.g., peaks and zero points), and improve the navigation accuracy in the test scenario by 34 \%. 

When using for motion detection, one main benefit of the AI method is that it can eliminate the threshold setting for various conditions, such as different devices and different user motion types. This study claims that AI can provide better zero-velocity detection under challenging motions such as crawling and ladder climbing; however, it does not have a detailed discussion on this point. Also, the test conditions in this study are too limited to reflect the algorithm performances in various scenarios. This limitation is common for many studies that use AI to enhance inertial sensing. 

The research work \cite{Gonzalez-Fiacchini-2018} uses the Gaussian process (GP), support vector machine (SVM), and nuclear ridge regression to detect robot sideslip. Different from many similar works, this research conducts detection based on a regression model, instead of a classification model. The classification model only outputs one of the limited numbers of modes, while the regression model has continuous output variables. In the results, the accuracies of all three regression models for sideslip detection were over 80 \%, which were close to the accuracy of the SVM classification. However, the computational load when using the regression model was over an order of magnitude higher than that of the classification model. Also, the computational load of the regression model based on GP, SVM, and nuclear ridge regression increased sequentially. The computational load is an important factor when using the AI method. 

For the use of environmental constraints, the research in \cite{Tong-X-2018} applies the hidden Markov process (HMM) to predict the magnetic-field interference, thereby automatically adjusting the weight of the magnetic update in the attitude and heading reference system (AHRS) algorithm, so as to obtain a more accurate heading estimate. However, this research only compares the performance of the AHRS with and without the HMM module, instead of comparing the contribution of HMM and other magnetic-interference prediction methods. Thus, it lacks a comprehensive evaluation of the HMM method compared to the traditional approaches.

\begin{figure}
           \centering
           \includegraphics[width=0.48 \textwidth]{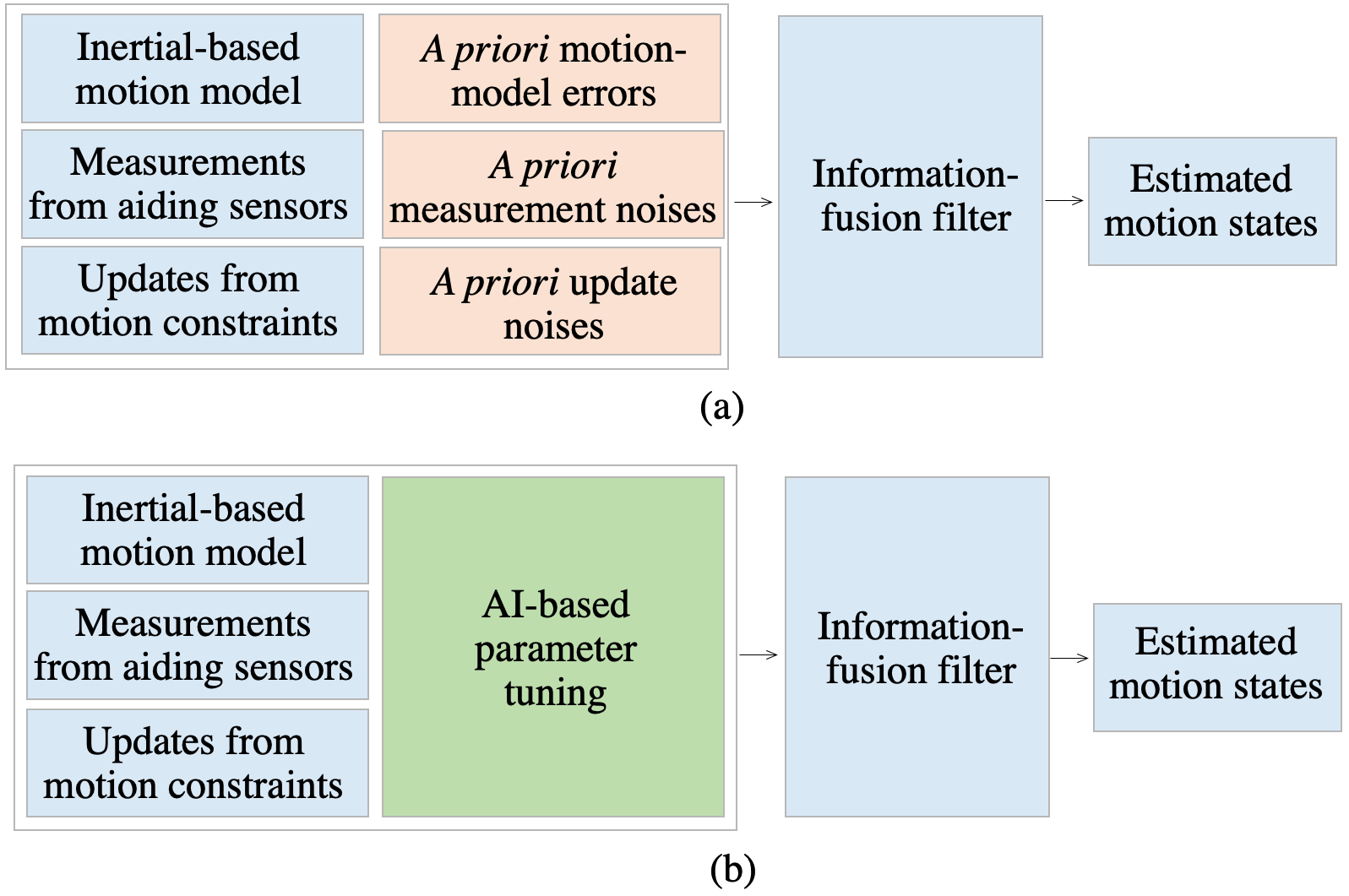}
           \caption{Multi-sensor information-fusion parameter tuning using traditional (a) and AI-based (b) methods}
           \label{fig:module-para-tuning}
         \end{figure}

When using a constraint, its uncertainty for different scenarios and various sensors is commonly adjusted through manual methods in practice \cite{Chen-Niu-2020}. If this is the case, \textit{a priori} information for the motion models and measurements are required in advance, as illustrated in Figure \ref{fig:module-para-tuning} (a). Such a requirement limits the scalability of the system. 

Therefore, researchers introduce ANN, adaptive filtering, and other methods to automatically perform adaptive multi-sensor data fusion \cite{Wu-Yang-2010} and motion-constraint uncertainty estimation \cite{Brossard-Barrau-2020}. As shown in  Figure \ref{fig:module-para-tuning} (b), the introduction of AI can mitigate or eliminate the requirement for \textit{a priori} information on motion models and measurements. The paper \cite{Wu-Yang-2010} uses an adaptive Kalman filter to determine the relative weight of multi-sensor information. Also, it designs a stepped adaptive Kalman filtering to control the effect of external measurement outliers. Meanwhile, the research in \cite{Brossard-Barrau-2020} uses ANN to adjust navigation filter parameters. It used gyros with bias stability of 36 deg/h to achieve the navigation accuracy similar to a light detection and ranging (LiDAR) odometry. The position error was 1.48 \% of the moving distance after 3.2 kilometers of driving. 

However, this paper has not theoretically explained why this performance can be achieved; thus, its performance in more scenarios needs further investigation. Also, the results can only show that ANN tuning led to more accurate navigation compared to the strategy without parameter tuning; however, they have not covered the comparison of ANN tuning and other parameter-tuning approaches. Meanwhile, the influence of the ANN algorithm structure and parameters on navigation results needs further study. 

In general, compared to manual operation and adaptive Kalman filtering, ANN has fewer assumptions and is theoretically more suitable for large-scale applications; however, it also puts forward higher requirements for the robustness of training data and algorithm.

To integrate with inertial sensors, a variety of sensors, such as GNSS \cite{Jaradat-Abdel-2014}, camera \cite{Milford-M-2013}\cite{Rambach-Tewari-2016}, odometer \cite{Li-Wang-2014}, and magnetometer \cite{Tong-X-2018}, have been involved. The paper \cite{Milford-M-2013} uses brain-inspired algorithms to process image data, which can provide position and attitude corrections for inertial motion tracking. Although brain-inspired algorithms are mostly used for visual navigation and simultaneous localization and mapping (SLAM), it may be directly used to process inertial data because it contains speed and orientation modules, which are consistent with the output of inertial systems. The research in \cite{Rambach-Tewari-2016} uses long short-term memory (LSTM) ANN enhanced INS to aid the visual odometry and reduce its mismatch rate from 7 - 15 \% to 0 \%. However, this study did not compare the impact of ANN- and geometric-based inertial solutions on visual odometry. In contrast, the paper \cite{Li-Wang-2014} uses the ANFIS to predict inertial navigation errors based on the odometer and inertial navigation states. Its principle is similar to that in \cite{Jaradat-Abdel-2014}. The use of fuzzy inference reduced navigation errors by 60 \%. Similar to many research works, the study did not further discuss the relationship between ANN structure and navigation performance. Thus, it is difficult to predict the system performance when the training and test data are collected in various scenarios. 

Most related works use AI to enhance data fusion in a loosely-coupled way, such as the fusion of GNSS and inertial-based position solutions. Meanwhile, there is a small amount of research using AI in tightly-coupled and ultra-tightly-coupled multi-sensor integration. Examples of tightly-coupling are the integration of inertial sensor and GNSS \cite{Wu-Yang-2010} and visual sensors \cite{Rambach-Tewari-2016}, while an example of ultra-tightly-coupling is that uses inertial sensors to aid GNSS receiver design \cite{Jwo-Chuang-2012}. Different from \cite{Jaradat-Abdel-2014}, the paper \cite{Jwo-Chuang-2012} applies AI to predict the corrections of the raw measurements (e.g., Doppler observations) instead of the corrections of the navigation states. Regardless of the integration modes, most research ideas are to use AI to maintain system performance when the external measurement is not available. 

In general, the use of AI in sensor fusion is flexible. It can enhance not only a single sensor but also the overall multi-sensor system. Therefore, it is expected that future research will not be limited to the above aspects. There will be new AI-based approaches to enhance the performance and intelligence of navigation and motion-sensing from different aspects.

\subsection{Artificial-Intelligence-Enhanced Inertial-Sensing System Evaluation}

The approaches for inertial system testing and evaluation include theoretical analysis, simulation testing, laboratory testing, semi-field testing, and field testing \cite{Niu-Wang-2016}. The scenarios of these methods are gradually approaching real-world applications, while their costs are gradually increasing. Thus, the choice of evaluation methods requires a compromise between the similarity to the real world and the cost. Although we have not found literature on using AI technology to enhance the performance testing and evaluation of inertial-sensing systems, we expect that AI technology will play an important role in this topic. 

For example, the selection of inertial sensors is important but very challenging. The importance is that if the sensors at a lower level can be used to meet the task requirements, generally at least one order of magnitude of hardware cost can be saved. On the other hand, it is challenging to model the relationship between the inertial sensor level and the system performance quantitatively. The reason for this phenomenon is that their relationship is affected by factors such as operating environments and vehicle dynamics, which are complex and difficult to predict \cite{Hong-Lee-2005}. In the future, it may be possible to use AI technology for sensor selection to alleviate this problem. To achieve this goal, in addition to suitable AI algorithms, the guarantee of data amount and reliability are key.

\subsection{Artificial-Intelligence-Enhanced Inertial Data and Application}

\begin{figure}
           \centering
           \includegraphics[width=0.44 \textwidth]{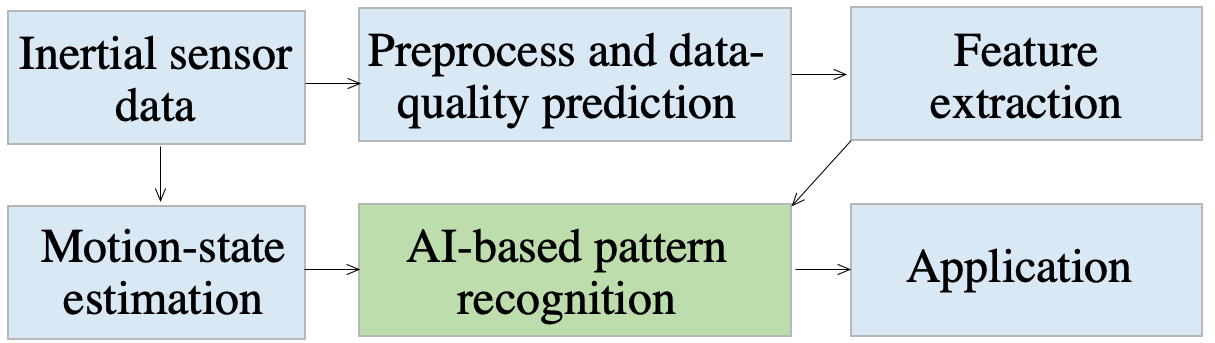}
           \caption{Flow chat of using inertial data for practical applications through AI-based pattern recognition}
           \label{fig:module-data-application}
         \end{figure}

The majority of the current AI-enhanced inertial sensing research works are on the use of data. Furthermore, among these works, the majority use AI to analyze the data features and implement pattern recognition, either classification or regression. Figure \ref{fig:module-data-application} demonstrates the process. The related applications involve not only positioning, navigation \cite{Jaradat-Abdel-2014}, and motion control \cite{Buskey-Wyeth-2001}, but also wearables, driving, big data, and inertial surveying. 

\textit{AI-enhanced motion control}: The paper \cite{Buskey-Wyeth-2001} makes a simple attempt to optimize ANN neurons in the application of inertial-based helicopter movement control. It discusses how to construct hidden layers with different numbers of neurons. In its test scenarios, 10 neurons provided more accurate solutions than 40 neurons. This result indicates that it is worthwhile to set the ANN structure properly according to the requirement, instead of using an ANN that is as complex as possible.

\textit{AI-enhanced wearables}: Related research has covered applications such as travel-pattern recognition \cite{Ravi-Wong-2016}, daily-activity recognition \cite{Moschetti-Fiorini-2016}, disease assessment \cite{Eskofier-B-2016}, gender and age recognition \cite{Khabir-K-2019}, the recognition and analysis of motions from workers \cite{Kim-Cho-2020} and athletes \cite{Windau-Itti-2019}, and personal safety \cite{Li-Xie-2019}. The research in \cite{Ravi-Wong-2016} applies ANN to identify walking, stationary, running, driving, and other motion types in multiple data sets and achieved an overall accuracy rate of over 95 \%. For real-time use on low-cost terminals, this study optimizes ANN calculation by adding frequency domain preprocessing and using a simplified ANN structure. 

The purpose of \cite{Moschetti-Fiorini-2016} is to verify the feasibility of using unsupervised algorithms to process IMU data at fingers and wrists for motion recognition. It uses unsupervised algorithms (SVM and random forests) and supervised algorithms (K-means clustering and another method) to recognize eight hand movements and achieved accuracy of over 90 \%. The unsupervised algorithms had higher recognition accuracy than the supervised ones in the tests. This outcome is not in line with expectations. However, the reason for this phenomenon has not been explained. 

The paper \cite{Eskofier-B-2016} concludes that the performance of ANN is superior to SVM, AdaBoost.M1, K-NN in IMU-based action classification based. Also, it points out that ANN relies less on feature selection compared to several other methods. However, the data of the study was limited. It included only three days of activity data from two Parkinson's patients. Meanwhile, the study points out that ANN is not as straightforward to explain the internal mechanism as several other methods. 

The research in \cite{Khabir-K-2019} attempts to use IMU data for gender and age identification, which is more challenging than motion recognition. The study used classification and regression models to process waist data of 744 people from the age of 2 to 78 when they walked and achieved a gender-recognition accuracy of 86 \%. In contrast, the age-recognition results have significant errors, especially for older people. The results of this research have a significant overfitting phenomenon. That is, the recognition accuracy is high on the training data but low in test data. Overfitting is an issue that needs to be solved in current AI-based research.

Also, the literature \cite{Kim-Cho-2020} uses ANN, random forests, and SVM to recognize 13 action types based on the data of 17 IMUs on the human body and achieved recognition accuracy of 82 \% to 94 \%. These methods are capable to recognize most of the selected action types accurately but meet a challenge in distinguishing a few movements (e.g., walking and squatting). Thus, additional information is needed. The study also explored the optimization of the number and installation location of IMUs for motion recognition purposes. 

Based on motion recognition, motion feedback is also considered in the research work \cite{Windau-Itti-2019}, which uses 13 IMUs on the human body and the recurrent neural network (RNN) algorithm to calculate the importance of each body part to the golf swing and then feedback which part needs to be improved. Through 100 swing training and AI feedback, the average score can be increased by 3.7 times. However, similar to many studies, this research only uses RNN as a toolkit and has not explored its internal mechanism.

\textit{AI-enhanced driving}: When using AI to enhance the use of inertial data, applications such as distraction detection \cite{Li-Xie-2019} and road condition detection \cite{Allouch-Kouba-2017} are involved. The research paper \cite{Allouch-Kouba-2017} uses inertial sensors in the smartphone, which is placed in a car, to monitor road conditions in real time. It uses three algorithms (C4.5 decision tree, SVM, and Naive Bayes) to identify the two types of road states, including smooth and uneven. The classification accuracies of all three methods were over 96 \% in the test scenario. However, the sample size of this study is relatively limited; only 2,000 samples from a smartphone were used. 

\textit{AI-enhanced motion big-data analysis}: The motion big data has been used for spatiotemporal analysis \cite{Gonzalez-Hidalgo-2008} and trend prediction \cite{Han-Liang-2015}. The paper \cite{Gonzalez-Hidalgo-2008} analyzes the movement of 100 thousand smartphone users for half a year and finds that the daily trajectory has a significant spatiotemporal pattern. For example, most activities are concentrated in a few locations. The regularity of the travel trajectory is helpful for applications such as epidemic prevention. If more dimensions of motion data are used, more phenomena may be discovered. The research in \cite{Han-Liang-2015} also shows that motion big data in smartphones can be used for trend prediction and environmental monitoring. However, this application has an important challenge in data preprocessing, such as compression and feature extraction. 

\begin{figure}
           \centering
           \includegraphics[width=0.48 \textwidth]{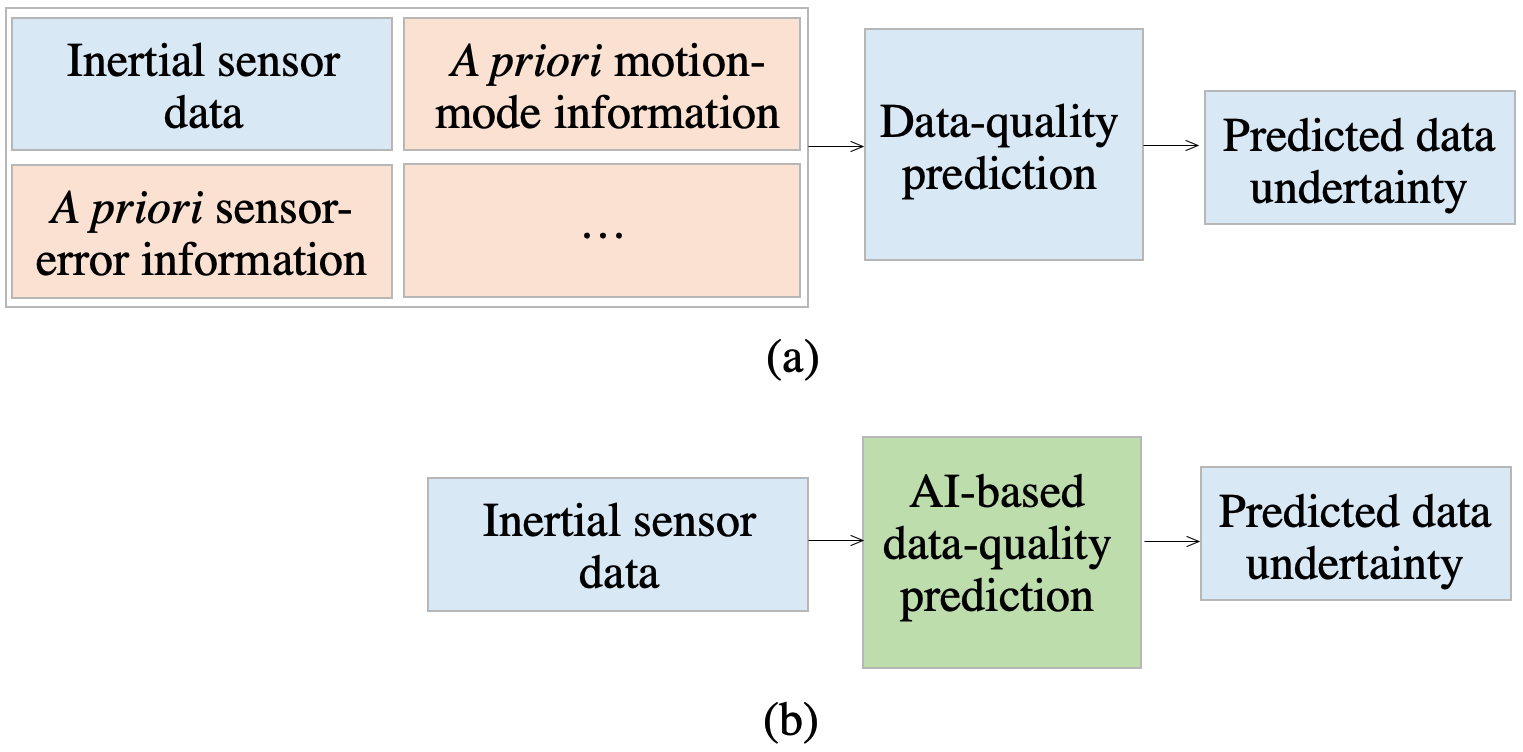}
           \caption{Inertial sensor data-quality prediction using traditional (a) and (b) based methods}
           \label{fig:module-data-quality}
         \end{figure}

Meanwhile, the paper \cite{Li-He-2018} points out that for motion big data, the acquisition of data quality indicators is important. The study proposes criteria for motion data-quality assessment, taking into account factors such as time length, motion mode, and sensor errors. Afterward, it automatically predicts data quality from the data, geometry, and database levels. Thus, \textit{a priori} information on factors such as motion mode and sensor error is required, as shown in Figure \ref{fig:module-data-quality} (a). 

To eliminate the need for such \textit{a priori} information, the AI-based method can be used, as demonstrates in Figure \ref{fig:module-data-quality} (b). The literature \cite{Li-Gao-2019} applies ANN to predict data quality. By comparing \cite{Li-He-2018} and \cite{Li-Gao-2019}, both geometric and AI methods can effectively predict the quality of motion data, so as to automatically select the most valuable motion data. However, these two approaches have not yet been combined effectively. Future research on this topic is needed.

\textit{AI-enhanced inertial surveying}: The related applications include pipeline \cite{Li-Cai-2016}, rail \cite{Lertxundi-Diez-2008}, gravity \cite{Li-X-2009} surveying, mobile mapping, and augmented reality (AR) \cite{Rambach-Tewari-2016}. Among them, the literature \cite{Li-X-2009} examines the improvement of ANN on land-vehicle-based INS/GNSS integrated gravity-disturbance measurement. When there was sufficient supervised training data, ANN improved the horizontal and vertical accuracies by over 400 \% and 100 \%, respectively. However, when the supervised training data is insufficient, ANN can still enhance the accuracy but the effect is significantly reduced. Meanwhile, this study points out that compared to Kalman filtering, the ANN-based approach still lacks complete theoretical support when using for gravity-disturbance measurement.

Table \ref{tab-literature} shows that the input of AI models in these applications is mostly inertial sensor data, and the output depends on the application requirements. Therefore, most of these applications can skip the need for geometric inertial algorithms. Also, most of these works apply classification algorithms, which are relatively mature in the current AI algorithms.

The existing research works have various vehicle platforms, including cars \cite{Allouch-Kouba-2017}, robots \cite{Gonzalez-Fiacchini-2018}, pipelines \cite{Li-Cai-2016}, railways \cite{Lertxundi-Diez-2008}, helicopters \cite{Buskey-Wyeth-2001}, wearables \cite{Ravi-Wong-2016}, etc. Among them, wearable applications have the most publications. For the use of wearables, there are multiple device locations, including the wrist (using smartwatches or bands) \cite{Ravi-Wong-2016}, feet (shoes) \cite{Wagstaff-Kelly-2018}, fingers (rings) \cite{Moschetti-Fiorini-2016}, head, waist, chest, arms, and legs \cite{Kim-Cho-2020}. The device (e.g., smartphones \cite{Chen-Zhao-2020}) may also be decoupled from the human body. The majority of applications need only one IMU, while human-motion analysis requires multiple IMUs [25, 26]. 

\begin{figure}
           \centering
           \includegraphics[width=0.48 \textwidth]{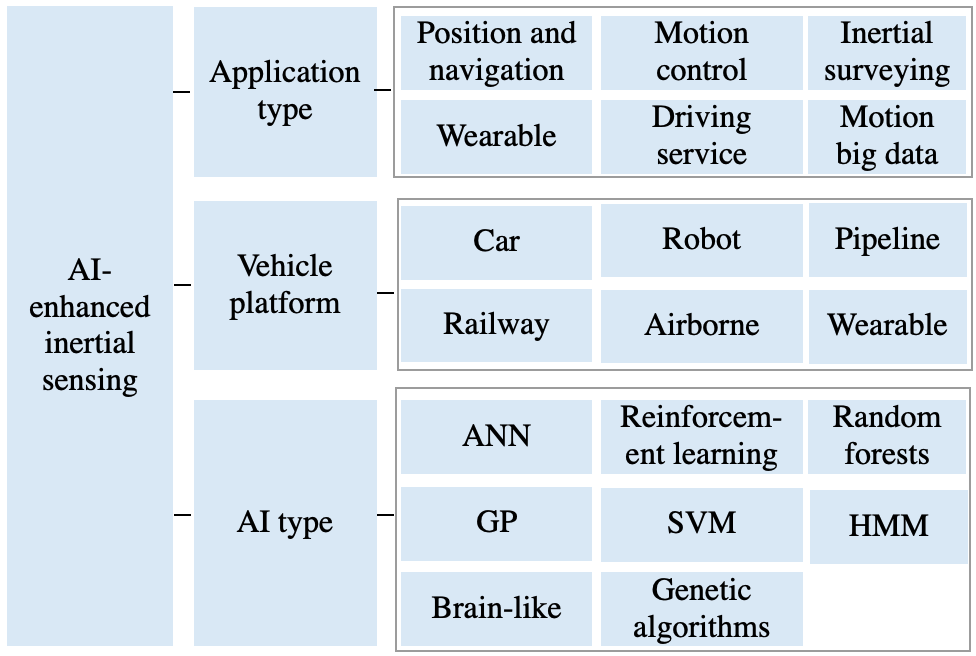}
           \caption{Existing AI-enhanced inertial sensing applications}
           \label{fig:class}
         \end{figure}

Figure \ref{fig:class} shows the classification of existing publications in terms of application type, vehicle platform, and AI type. These studies can provide theoretical and methodological guidance for other applications that use AI for data processing. With the miniaturization and technological advancement of inertial sensors, they will become more ubiquitous in motion-sensing applications. Correspondingly, the application of AI technology will also be increasingly diverse.        
         
\section{Artificial-Intelligence Algorithms in Inertial Sensing}
 \label{sec-AI-algorithms}
 
\textcolor[rgb]{0,0,0}{ 
As shown in Table \ref{tab-literature}, various AI algorithms have been applied for inertial sensing. Among these algorithms, ANN has been most widely used. In addition, there are other AI algorithms, such as deep reinforcement learning (DRL), GP \cite{Gonzalez-Fiacchini-2018}, random forest \cite{Adusumilli-Bhatt-2013}, SVM \cite{Yang-Ahn-2016}, HMM \cite{Tong-X-2018}, genetic algorithms \cite{Chong-S-2016}, and brain-inspired algorithms \cite{Milford-M-2013}. This section introduces the ANN, DRL, random forest, and GP algorithms. These four algorithms represent AI methods based on neural networks, reinforcement learning, decision trees, and random processes.
}

\subsection{Artificial Neural Networks}

\begin{figure}
           \centering
           \includegraphics[width=0.43 \textwidth]{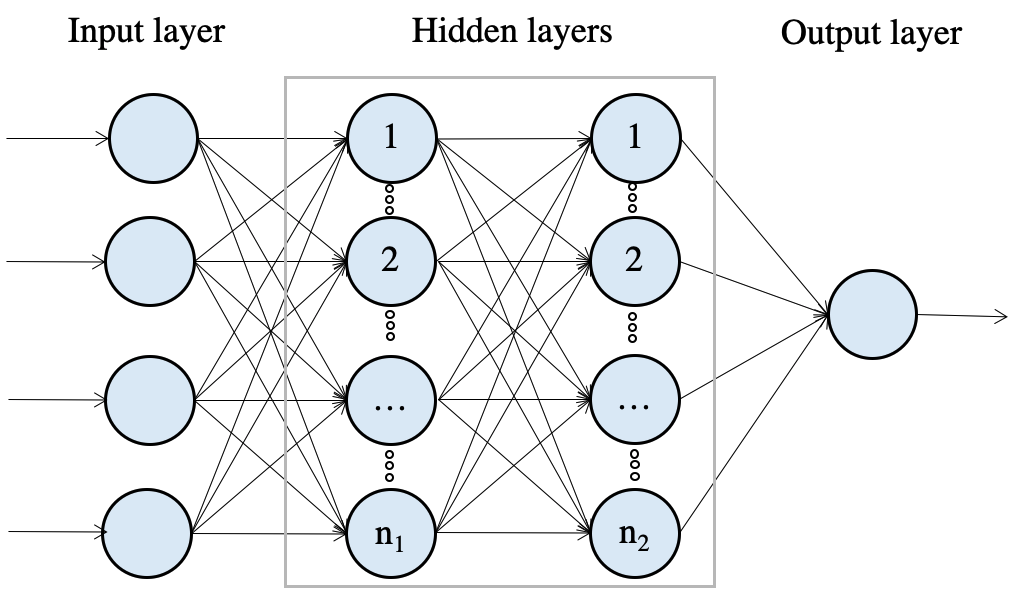}
           \caption{Typical artificial neural network algorithm architecture }
           \label{fig:ann}
         \end{figure}

\textcolor[rgb]{0,0,0}{ 
ANN is a computational model that mimics the structure and function of biological neural networks. It is commonly used to estimate or replace complex and unknown functions \cite{Rojas-R-2013}. An ANN implementation includes one input layer, one output layer, and at least one hidden layer. Each layer contains at least one neuron. Figure \ref{fig:ann} shows a schematic diagram of the ANN structure. As summarized in Table \ref{tab-literature}, when used for inertial sensing, the input layer data is commonly inertial sensor data [11, 12, 15, 22, 25], data features [21, 29], navigation status [9, 10, 18, 19], and environmental information [5, 39]. The output layer is the result of pattern recognition [12, 20-22, 24, 25] or state estimation [9, 11, 15]. There are mostly 1 to 3 hidden layers; the number of neurons in each layer varies from 1 to 1000 orders of magnitude. 
}

\textcolor[rgb]{0,0,0}{ 
When using an ANN, it is key to estimate the values of neuron parameters (e.g., weights and biases) to obtain an expected output based on a given input. To achieve this goal, approaches such as back-propagation \cite{Hagan-Menhaj-1994} are commonly applied. Among the types of ANNs, LSTM ANN \cite{Chen-Zhao-2020}, convolutional neural network (CNN) \cite{Brossard-Barrau-2020}, and ANFIS \cite{Jaradat-Abdel-2014} are relatively widely used in the existing inertial-sensing works. LSTM and CNN are mostly used to deal with pattern-recognition problems [12, 21, 28]. Also, they are used for state estimation [15, 19]. In contrast, ANFIS is mostly used to correct navigation results \cite{Jaradat-Abdel-2014}. Meanwhile, there are other types of ANNs, such as radial} basis function (RBF) ANN \cite{Li-Ding-2010}, which have been used for gyro bias estimation in several papers.

Summarizing the relevant literature, ANN has the following advantages: (1) it has been widely used because it imposes few restrictions on the input variables. For example, it does not require \textit{a priori} information on the distribution of sensor errors \cite{Chen-X-2003}. Also, it does not rely on the selection of features as other methods \cite{Eskofier-B-2016}. (2) The ANN algorithm has been well-developed. It has been successfully applied in many fields, such as image processing and speech recognition. (3) It is straightforward to implement. Various open-source algorithm platforms and tools can be utilized directly. 

The challenges for the use of ANN in inertial sensing include: (1) the ANN algorithm has been widely used as a black box or gray box. Few of related research can provide an explicit model expression of its working principle. Thus, it is difficult to optimize the internal algorithm, especially when the model is complex. (2) The ANN structure (e.g., the number of hidden layers and the number of neurons) directly affects the classification or estimation result. However, there is not enough theoretical guidance for the setting of the ANN structure. Thus, most research works can only optimize the ANN structure based on data processing results. The obtained ANN structure is difficult to handle complex or new navigation scenarios. (3) To ensure performance, a large number of neurons are usually needed. Accordingly, a large amount of training data is required. However, the amount of data in most existing literature cannot meet the needs of training a robust neural network. Crowdsourcing has great potential in providing big data; however, its data quality is difficult to guarantee \cite{Li-He-2018}.

\subsection{Reinforcement Learning}

\begin{figure}
           \centering
           \includegraphics[width=0.36 \textwidth]{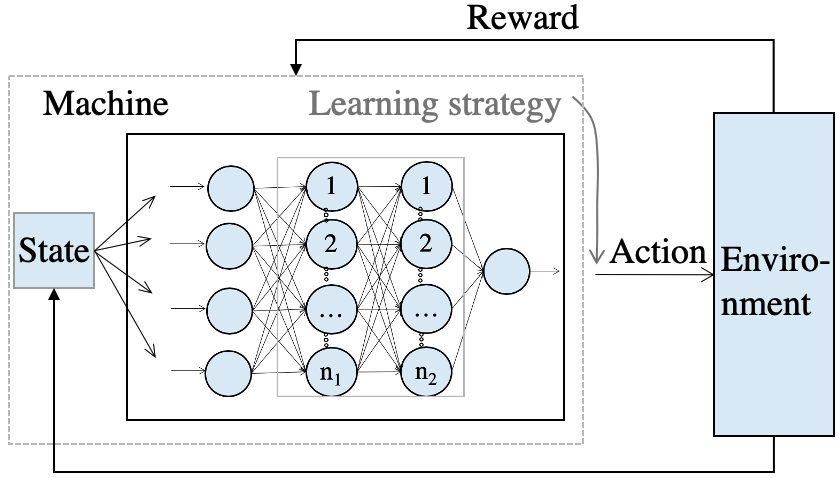}
           \caption{Typical deep reinforcement learning algorithm architecture }
           \label{fig:drl}
         \end{figure}

DRL \cite{Hessel-M-2017} is a combination of deep learning and reinforcement learning. Reinforcement learning provides a learning goal for deep learning. The basic idea of reinforcement learning is to gradually learn the optimal learning strategy to complete the goal by maximizing the cumulative reward from the environment. Figure \ref{fig:drl} shows a schematic diagram of the DRL algorithm structure. In the existing literature, DRL has been successfully applied in path planning navigation in mazes and action games. Also, it has begun to be applied in real-world positioning \cite{Li-Hu-2019}. The DRL is used wireless positioning using unsupervised training data, thereby reducing the need for manual intervention. On the other hand, the research shows that DRL-based wireless positioning accuracy is still difficult to reach that from supervised learning. There are challenges such as the requirement of high data and computational loads and the difficulty in setting reward models.
         
In the inertial-sensing field, the research papers \cite{Nobre-Heckman-2019} and \cite{Hu-Liu-2018} use DRL and reinforcement learning for calibration-action optimization and navigation filter-parameter tuning, respectively. In particular, the paper \cite{Hu-Liu-2018} uses reinforcement learning to adjust the multi-sensor integrated navigation Kalman filter parameters (e.g., the errors in GNSS positions and inertial sensor outputs) stepwise in increments of +1, -1, 0. Compared to brute-force-based parameter tuning, the reinforcement-learning-based method had less computational load. However, when the number of parameters increases, it becomes challenging to construct an appropriate Markov decision process to adjust all parameters in a stepwise manner.
         
The advantages of DRL are as follows. (1) Compared to ANN, DRL provides extra decision-making ability because of the introduction of reinforcement learning. Therefore, DRL is suitable for control and decision-making problems, such as path planning and action optimization. (2) In addition to state estimation, DRL considers a long-term reward mechanism. Thus, it does not need to recalculate the optimization strategy when the external environment changes \cite{Hu-Liu-2018}. 

The challenges of using DRL in inertial sensing include: (1) The DRL performance is directly affected by the setting of rewards. However, the theoretical model of reward setting is difficult to obtain; how to set rewards reasonably remains a challenging issue. (2) As a combination of deep learning and reinforcement learning, DRL has a heavy computational load. To accelerate computation, the support from future neural network chips \cite{Pei-J-2019} is expected.

\subsection{Random Forests}
          \begin{figure}
           \centering
           \includegraphics[width=0.36 \textwidth]{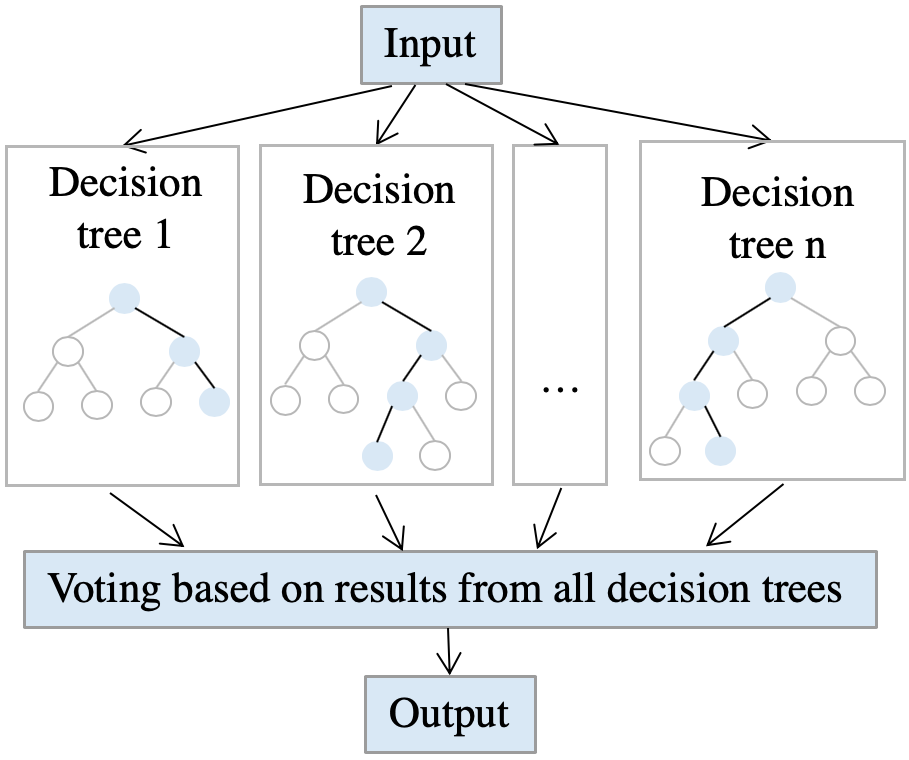}
           \caption{Typical random forests algorithm architecture}
           \label{fig:rf}
         \end{figure}  
         
Random forests \cite{Breiman-L-2001}, which is a decision-tree-based method, is implemented in three steps: sub-sampling, decision-tree training, and prediction. In the sub-sampling step, a fixed number of sub-samples are randomly selected, and each sub-sample contains a fixed number of randomly-selected features. Afterward, the selected sub-samples are trained in the decision-tree training step. In the prediction phase, all decision trees vote, and the voting results are summarized to generate estimation results. Figure \ref{fig:rf} shows a schematic diagram of the random-forest algorithm. The research paper \cite{Adusumilli-Bhatt-2013} predicts changes in target position by inputting motion information such as velocity and heading angle into random forests.

The advantages of random forests include: (1) Compared to the ANN method, the random-forest model is more straightforward to interpret; also, the parameters are simpler to set. (2) Random forests can output not only the estimated results but also the relative weight of the results. 

The challenges of using random forests include: (1) When a large number of decision trees are used to handle a complex task, the calculation efficiency is low. (2) Random forests may be overfitting and sensitive to noise \cite{Briem-Benediktsson-2002}.

\subsection{Gaussian Processes}

\begin{figure}
           \centering
           \includegraphics[width=0.42 \textwidth]{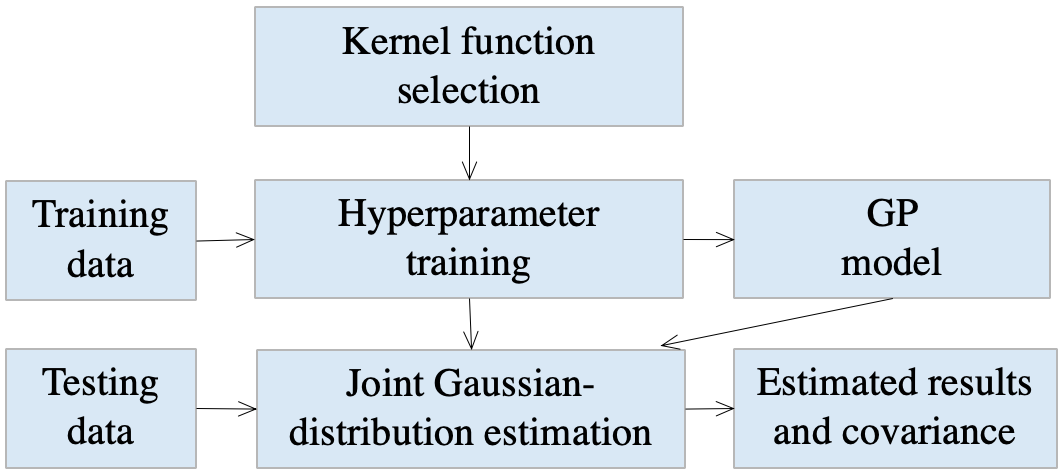}
           \caption{Typical Gaussian process algorithm architecture }
           \label{fig:gp}
         \end{figure}

GP is an estimation method for signal distribution in a continuous domain (e.g., a time or space domain) \cite{Rasmussen-C-2003}. Figure \ref{fig:gp} shows a schematic diagram of the GP algorithm. GP can be understood as the union of a series of continuous random variables. Meanwhile, the random variables at each time or space point follow Gaussian distribution. Thus, the GP can reflect the correlation between different observations. 

The GP model can be uniquely determined by the mean function and the kernel function (i.e., covariance function). The mean function is usually zeroed in advance or determined with a geometric model. There are various kernel functions, such as RBF kernel, exponentiated quadratic kernel, and rational quadratic kernel \cite{Rasmussen-C-2003}. The research paper \cite{Gonzalez-Fiacchini-2018} uses GP to detect the side slip of robots. The GP-based regression model achieved a prediction accuracy similar to SVM classification. Also, according to the results, the computation load for GP regression was less than SVM, and nuclear ridge regression.

The advantages of using GP in inertial sensing include: (1) The GP model can be physically explained and visualized. This characteristic makes it associated with the spatial distribution of data. Meanwhile, it might be easier to be applied together with a geometric algorithm. (2) Besides the estimated results, the GP can output their uncertainties, which is important when using the results by machines. (3) Compared to other AI methods, GP has fewer parameters in the model, which makes it possible to be trained by using less training data; also, it is straightforward to be applied.

The challenges of using GP include: (1) It has a Gaussian-process assumption. Although a Gaussian model can be used to approximately model many inertial-sensing signals, the existence of systematic errors in the signals may degrade the GP performance. (2) Because a GP model has limited parameters, it may be difficult to model as many details as other complex models (e.g., ANN) when there are sufficient training data. 

In general, AI methods based on neural networks, reinforcement learning, decision trees, and random processes have their advantages and challenges, as shown in Table \ref{tab-ai-methods}. Various methods are more suitable for solving different problems. For example, from the existing literature, ANN has been widely applied to enhancing all inertial-sensing steps. In contrast, DRL has been used more on control and decision-making issues. Random forests are used for classification problems, while GP is used in data interpolation. If different types of AI methods can be effectively combined, a better AI model may be obtained to further enhance inertial sensing.

\begin{table*}
           \centering
\begin{tabular}{p{2.2cm} p{6.4cm} p{6.4cm}}
\hline
\textbf{AI algorithm type} & \textbf{~~~~~~~~~~~~~~~~~~~~~~~~Advantage}  &  \textbf{~~~~~~~~~~~~~~~~~~~~~~~~Challenge}   \\ \hline
ANN 
& 
\begin{itemize}
  \item	Imposes few restrictions on input
  \item	Well-developed and most widely used
  \item	Abundant resources and open-source algorithm platforms and tools
\end{itemize}
	& 
\begin{itemize}
  \item	Black/grey box issue
  \item	Lacks sufficient theoretical guidance on network structure
  \item	Require a large amount of data to train a complex network; heavy computational load
\end{itemize}
	\\ 
Reinforcement learning (e.g., DRL) 
& 
\begin{itemize}
 \item	Has extra decision-making ability; suitable for control and decision-making problems
 \item	Considers long-term reward; does not need to recalculate optimization strategy when external environment changes
\end{itemize}
	& 
\begin{itemize}
 \item	Theoretical model of reward setting is difficult to obtain
 \item	Require a large amount of data to train a complex network; a heavy computational load
 \item	Black/grey box issue
\end{itemize}
	\\ 
Decision tree based (e.g., random forests) 
& 
\begin{itemize}
  \item	More straightforward for interpretation and parameter setting
  \item	Can output relative weight of estimated results
\end{itemize}
	& 
\begin{itemize}
  \item	Calculation efficiency is low when a large number of decision trees are used to handle a complex task
  \item	May be overfitting and sensitive to noise
\end{itemize}
	\\ 
Random process based (e.g., GP) 
& 
\begin{itemize}
  \item	Model can be physically explained and is associated with the spatial distribution of data 
  \item	Can output uncertainties of estimated results
  \item	Has fewer parameters in the model; can be trained by using less training data
\end{itemize}
	& 
\begin{itemize}
  \item	Limited by the Gaussian-process assumption
  \item	Difficult to model as many details as other complex models when there are sufficient training data
\end{itemize}
	\\ 
 \hline
           \end{tabular}
           \caption{ Advantages and challenges for main AI methods in inertial sensing   }
           \label{tab-ai-methods}
         \end{table*}
    
\section{Benefits and Challenges of Artificial-Intelligence-Based Inertial Sensing}
\label{sec-pros-and-cons} 

This section analyzes the advantages and challenges of using AI for inertial sensing. Some possible solutions for the challenges are also provided.    

\subsection{Benefits}

To achieve high motion-sensing performance, accurate modeling and parameter setting are usually required in advance. However, this premise is difficult to meet in some applications. For example, inertial-sensing performance is affected by factors such as sensor errors, vehicle dynamics, and application scenarios. The influence of these factors is complicated and difficult to predict, making it difficult to obtain accurate models and parameters in advance. Therefore, manual intervention is required to model and tune parameters for specific scenarios. Using AI technology can alleviate this problem. By summarizing existing literature, AI technology can bring the following benefits in inertial sensing:

\begin{enumerate}
\item AI can be applied to enhance parameter estimation. For example, it can accelerate the convergence of a Kalman filter \cite{Chiang-Huang-2010} and improve estimation accuracy \cite{Nobre-Heckman-2019} \cite{Brossard-Barrau-2020}. The combination of AI and traditional time-series analysis methods \cite{Kim-Agrawal-2016}, geometric motion models \cite{Chiang-Huang-2010} \cite{Jaradat-Abdel-2014}, and data quality analysis methods \cite{Li-He-2018} can achieve better performance.

\item  AI can be used to replace complex models, especially for scenarios that have data with strong nonlinearity, strong correlation, and high-order stochastic errors. For example, it may be used to replace part or all of the mathematical models between inertial sensor data and motion states \cite{Chen-Zhao-2020}, alleviating error sources such as initial-alignment errors and misalignment errors. Furthermore, using AI may alleviate some inherent problems of geometric dead-reckoning models, such as the accumulation of navigation errors.

\item AI can be utilized for adaptive parameter tuning. The introduction of AI can eliminate the threshold setting for different devices and various motion modes \cite{Wagstaff-Kelly-2018}. Also, AI can enhance the adaptive weighting of multi-sensor information [14, 16] and prediction of motion-constraint uncertainty \cite{Brossard-Barrau-2020}.

\item The use of AI can reduce the requirements of \textit{a priori} information, such as the sensor error distribution \cite{Chen-X-2003} and the selection of features \cite{Eskofier-B-2016}. This characteristic may enhance the scalability of the inertial-sensing system.

\item AI can be applied to enhance classification. The current AI method has obtained accurate results on the classification of actions \cite{Ravi-Wong-2016} \cite{Moschetti-Fiorini-2016} \cite{Kim-Cho-2020}, gender \cite{Khabir-K-2019}, and road conditions \cite{Allouch-Kouba-2017} based on inertial sensor data. Some applications are even difficult to complete without using AI methods.

\item AI can be utilized to strengthen control and decision-making. Using AI to replace manual control \cite{Hsu-C-2011} and decision-making \cite{Nobre-Heckman-2019} modules is conducive to the realization of autonomous systems; also, it is conducive to the completion of professional tasks by people without relevant professional experiences \cite{Nobre-Heckman-2019}. Also, AI-based motion feedback can effectively improve motion optimization \cite{Windau-Itti-2019}.

\item AI can enhance the testing and evaluation of inertial systems. For example, the use of AI technology may alleviate the challenge of quantifying the relationship between navigation performance and factors such as sensor errors, vehicle dynamics, and operating environments. This relationship is key to achieve appropriate sensor selection and system performance prediction in advance.

\item The AI-based inertial-motion-tracking module can be used to assist other sensors, such as reducing the mismatch rates of visual odometry \cite{Rambach-Tewari-2016} and predicting the correction of GNSS measurements \cite{Jwo-Chuang-2012}.

\item The use of unsupervised AI methods can eliminate the need for labeled training data, which is essential for traditional methods but time-costly and labor-intensive. Then, machines can perform parameter estimation (such as localization \cite{Li-Hu-2019}) and pattern recognition \cite{Moschetti-Fiorini-2016} automatically.

\end{enumerate}

\subsection{Challenges}

According to existing literature, using AI in inertial sensing also has challenges, including:

\begin{table*}
           \centering
\begin{tabular}{p{1.8 cm} p{14.5cm}}
\hline
\textbf{Advantages} 
& 
\begin{itemize}
 \item	Can enhance parameter estimation. Also, the combination of AI and traditional methods can achieve better performance
 \item	Can replace complex models in challenging scenarios. Furthermore, it may alleviate some inherent problems of geometric dead-reckoning models
 \item	Can be utilized for adaptive parameter tuning and reduce manual intervention
 \item	Can reduce the requirements of a prior information
 \item	Can enhance classification
 \item	Can strengthen control and decision-making
 \item	Can enhance the testing and evaluation of inertial systems
 \item	Can further assist other sensors
Can eliminate the need for labeled training data
\end{itemize}
	\\ 
\textbf{Challenges} 
& 
\begin{itemize}
 \item	Lacks explicit mathematical models like traditional methods and thus depends more on data amount and quality. It is difficult to control performance degradation when training data is insufficient or not labeled properly. Also, some problems are difficult to solve at the data level
 \item	The overfitting problem. A lack of cross-validation in most research works
 \item	The black/grey box issue. A lack of a systematical theoretical framework in the AI structure and parameter setting
 \item	May mask the physical and geometric phenomena behind the problem 
 \item	May lose the short-term accuracy and reliability of the geometric inertial algorithm
 \item	Has a much higher requirement computing power compared to geometric methods
 \item	Most relevant works use supervised AI algorithms, which require labeled training data
 \item	Most relevant works focus on classification, instead of parameter estimation. Many aspects of inertial sensing have not been enhanced 
 \item	A lack of public datasets and evaluation frameworks
\end{itemize}
	\\ 
\textbf{Prospects} 
& 
\begin{itemize}
 \item	Solve data problems
 \item	Cognition and optimization of internal mechanisms of complex AI structures
 \item	AI-based parameter estimation
 \item	Combination with physical and geometric principles
 \item	The use of unsupervised AI methods
\end{itemize}
	\\ 
\hline
           \end{tabular}
           \caption{ Benefits, challenges, and prospects for using AI in inertial sensing  }
           \label{tab-ai-in-inertial-sensing}
         \end{table*}

\begin{enumerate}
\item Most of the existing AI algorithms are data-driven, lacking explicit mathematical models like traditional methods such as Fourier analysis and Kalman filtering. Thus, sufficient and reliable data is a necessary condition for training an accurate AI model. However, it is difficult to obtain a large amount of data in professional inertial-sensing applications; thus, most of the related research works do not have sufficient training data for robust AI models. Also, for mass-market applications such as smartphone-based crowdsourcing, it is challenging to select a few reliable motion data from massive crowdsourced data \cite{Li-He-2018} and to preprocess massive data reliably \cite{Han-Liang-2015}. In addition, it is important but challenging to control the performance degradation of AI methods when there is insufficient training data \cite{Li-X-2009} or when the data has not been labeled properly. Meanwhile, some problems are difficult to solve at the data level, such as the recognition of actions that have similar motion features \cite{Kim-Cho-2020}.

\item Many of the existing research works suffered from the overfitting problem. Their results can verify the potential of using ANN to enhance inertial sensing \cite{Jaradat-Abdel-2014}. However, they cannot guarantee how much performance improvement can be achieved. The reason for this phenomenon is that the training and testing scenarios are limited; thus, there is a lack of cross-validation in various scenes. In other words, it is difficult for users to predict how accurate the system can achieve if test scenario changes. To alleviate this problem, it is necessary to increase the data amount and diversity, adjust the sample-selection method, and use additional data for cross-validation.

\item When using AI to enhance inertial sensing, a black/grey box issue occurs. To improve performance, the current papers often use complex AI structures, such as an ANN with a large number of layers and neurons \cite{Rambach-Tewari-2016}. If this is the case, they can only set AI parameters through attempts. It is difficult to systematically optimize the internal data-processing mechanism, such as the ANN structure and parameters, the number and depth of decision trees, and the setting of reward in reinforcement learning. The paper \cite{Buskey-Wyeth-2001} has compared the performance of various numbers of neurons on helicopter control. Also, it shows that more complex AI models not always lead to better performance. However, this research work lacks a systematical theoretical framework in parameter setting. Investigating the internal AI mechanism is key to its optimization and scalability.
Furthermore, the current research has not fully explored the mapping relationship between the AI algorithm and the performance of specific tasks, such as initial alignment \cite{Chiang-Huang-2010}, navigation \cite{Brossard-Barrau-2020}\cite{Li-Wang-2014}, and pattern recognition \cite{Windau-Itti-2019}. This missing of such a relationship limits the significance of the work to peers.

\item When using AI to replace geometric algorithms, the physical and geometric phenomena behind the problem may be masked. For example, current research on AI-based inertial sensing rarely explores the relationship sensing performance and factors such as sensor errors, vehicle dynamics, and operating environment. As a result, it is difficult to predict the system performance before the task or to plan sensor selection and trajectory design. To alleviate this problem, the relationship between these factors and data quality is needed. For this purpose, it may be useful to combine with traditional methods, such as observability analysis. 

\item When using AI to replace geometric algorithms, the inertial-sensing system may lose the short-term accuracy and reliability of the geometric inertial algorithm, which is a key advantage of inertial-based motion sensing. This phenomenon can be seen from the results in \cite{Chen-Zhao-2020}. To alleviate this issue, it is recommended to fuse AI with geometric algorithms. For a profound fusion, how to determine the weights of geometric and AI algorithms in different scenarios is a challenge.

\item AI-based inertial sensing has a much higher requirement computational load compared to geometric methods. Moreover, using regression models for parameter estimation suffers from a heavier computational load than the classification problem \cite{Gonzalez-Fiacchini-2018}. The development of AI chips, computing technologies (e.g., edge computing and fog computing), and 5G data transmission technologies may alleviate this problem. To meet the needs of real-time operation on low-cost terminals, it is still necessary to optimize ANN calculations \cite{Ravi-Wong-2016} by using techniques such as data preprocessing and simplified ANN structures.

\item Most of the existing related papers apply supervised AI algorithms, which require labeled training data. However, the data-labeling process is time-consuming and labor-intensive. Thus, it is important to use unsupervised AI methods to handle inertial-sensing problems. However, on many motion-sensing problems, the performance of unsupervised AI methods is still difficult to reach the level of supervised ones \cite{Li-Hu-2019}.

\item Most of the relevant work has focused on using inertial data for pattern classification; however, less is on the estimation of inertial-sensing models and parameters. A reason for this phenomenon is that the current AI algorithm is relatively stronger in classification. Many aspects of inertial sensing have not been enhanced with AI. For example, there are few publications on using AI technology to enhance inertial sensor hardware.

\item There is a lack of public datasets, which are key to the rapid development of AI research in recent years. This fact makes it difficult to compare the solutions from various publications. 

\end{enumerate}

\section{Conclusions}
 \label{sec-conclusion}
         
This article reviews the research on using AI technology to enhance inertial sensing from various aspects. In terms of classification, AI algorithms can already complete tasks independently without relying on traditional geometric models. However, in terms of positioning and navigation state estimation, most of the current research still needs a combination of traditional geometric algorithms and AI. Traditional algorithms are mostly based on theoretical mathematical models, which are rigorous but sometimes difficult to derive. In contrast, AI can enhance inertial-sensing in scenarios where the model is complex (such as nonlinear, strong correlation, high-order noise), varying (such as time-varying, spatial irregularities), difficult to predict, and difficult to build. When there is sufficient and reliable training data, the introduction of AI algorithms can provide more accurate inertial-sensing results with less manual intervention.
 
On the other hand, AI-enhanced inertial sensing faces challenges such as the dependency on reliable big data, the limited knowledge of complex AI structures, the limitations on using AI for parameter-estimation problems, and the limitation in quantifying the impact of AI on inertial sensing aspects. These challenges also reflect the opportunity for the combination of future AI and inertial sensing. Table \ref{tab-ai-in-inertial-sensing} lists the advantages, challenges, and prospects of using AI technology in inertial sensing. It is expected that with the further development of big data technology, AI algorithms, computing power, and AI chips, future AI technologies will be used more deeply and effectively for inertial sensing. In addition to traditional fields such as positioning, navigation, motion control, and mobile mapping, new fields such as new sensors, autonomous navigation, wearable computing, and motion big data, will be involved.

\clearpage

\end{document}